
\magnification\magstep 1
\def\Ai{\hbox{\hbox{${\cal A}$}}\kern-1.9mm{\hbox{${/}$}}}
\def\Vi{\hbox{\hbox{${\cal V}$}}\kern-1.9mm{\hbox{${/}$}}}
\def\Di{\hbox{\hbox{${\cal D}$}}\kern-1.9mm{\hbox{${/}$}}}
\def\lam{\hbox{\hbox{${\lambda}$}}\kern-1.6mm{\hbox{${/}$}}}
\def\D{\hbox{\hbox{${D}$}}\kern-1.9mm{\hbox{${/}$}}}
\def\A{\hbox{\hbox{${A}$}}\kern-1.8mm{\hbox{${/}$}}}
\def\pom{\hbox{I$\!$P}}
\def\V{\hbox{\hbox{${V}$}}\kern-1.9mm{\hbox{${/}$}}}
\def\parz{\hbox{\hbox{${\partial}$}}\kern-1.7mm{\hbox{${/}$}}}
\def\B{\hbox{\hbox{${B}$}}\kern-1.7mm{\hbox{${/}$}}}
\def\R{\hbox{\hbox{${R}$}}\kern-1.7mm{\hbox{${/}$}}}
\def\si{\hbox{\hbox{${\xi}$}}\kern-1.7mm{\hbox{${/}$}}}
\font\chap= cmbx10 scaled 1200
\font\ninerm= cmr9 scaled 1000
\font\itf= cmsl9 scaled 1000

\pageno=0

\line{
\vbox{\hsize 3.5 true cm
          \noindent }\hfill
\vbox{\hsize 4.0 true cm
           \noindent \bf LYCEN/95-37\break
                         November 1995}}

\vskip 2.0 true cm

\centerline{\bf THE POMERON IN ELASTIC AND DEEP INELASTIC SCATTERING}

\vskip 1.5  true cm

\centerline{ M. Bertini$^a$,  M. Giffon$^a$, L. Jenkovszky$^b$,
F. Paccanoni$^c$, E. Predazzi$^d$}

\bigskip
\bigskip

\centerline{ $^a$Institut de Physique Nucl\'eaire de Lyon, IN2P3-CNRS et
Universit\'e Claude Bernard, }

\centerline{43 Bld du 11 Novembre 1918 F-69622 Villeubanne Cedex, France}

\bigskip

\centerline{$^b$Bogoliubov Institut for Theoretical Physics,}

\centerline{Academy of Sciences of the Ukrain}

\centerline{252143 Kiev, Ukrain}

\bigskip

\centerline{$^c$ Dipartimento di Fisica, Universit\`a di Padova,}

\centerline{INFN Sezione di Padova, via F. Marzolo I-35131 Padova, Italy}

\bigskip
\centerline {$^d$Dipartimento di Fisica Teorica, Universit\`a di Torino, Italy}

\centerline{and INFN, Sezione di Torino, 10125 Torino, Italy}

\vskip 2.0 true cm
\centerline{\bf Abstract}
\medskip
We discuss some properties of the Pomeron in high energy elastic
hadron-hadron and deep inelastic lepton-hadron scattering. A number of issues
concerning the nature and the origin of the Pomeron are briefly recalled here.
The novelty in this paper resides essentially in its presentation; we strive at
discussing all these various issues in the following unifying perspective : it
is our contention that the Pomeron is one and the same in all reactions.
Various examples will be provided illustrating why we do not believe that one
should invoke additional tools to describe the data.
\vfill\eject
\vglue 1 true cm

\centerline{\bf THE POMERON IN ELASTIC AND DEEP INELASTIC SCATTERING}

\vskip 1.5  true cm

\centerline{ M. Bertini$^a$,  M. Giffon$^a$, L. Jenkovszky$^b$,
F. Paccanoni$^c$, E. Predazzi$^d$}

\bigskip
\bigskip

\centerline{ $^a$Institut de Physique Nucl\'eaire de Lyon, IN2P3-CNRS et
Universit\'e Claude Bernard, }

\centerline{43 Bld du 11 Novembre 1918 F-69622 Villeubanne Cedex, France}

\bigskip

\centerline{$^b$Bogoliubov Institut for Theoretical Physics,}

\centerline{Academy of Sciences of the Ukrain}

\centerline{252143 Kiev, Ukrain}

\bigskip

\centerline{$^c$ Dipartimento di Fisica, Universit\`a di Padova,}

\centerline{INFN Sezione di Padova, via F. Marzolo I-35131 Padova, Italy}

\bigskip
\centerline {$^d$Dipartimento di Fisica Teorica, Universit\`a di Torino, Italy}

\centerline{and INFN, Sezione di Torino, 10125 Torino, Italy}

\vskip 2.0 true cm
\centerline{\bf Abstract}
\medskip

\hsize=16 true cm
\leftskip= 1. true cm

We discuss some properties of the Pomeron in high energy elastic
hadron-hadron and deep inelastic lepton-hadron scattering. A number of issues
concerning the nature and the origin of the Pomeron are briefly recalled here.
The novelty in this paper resides essentially in its presentation; we strive at
discussing all these various issues in the following unifying perspective : it
is our contention that the Pomeron is one and the same in all reactions.
Various examples will be provided illustrating why we do not believe that one
should invoke additional tools to describe the data. For pedagogical
convenience, we list below the topics to be covered in the following.
\medskip

\hskip 1.5  true cm 1. Introduction. How many Pomerons?

\hskip 1.5  true cm 2. The Pomeron in the $S$-matrix theory

\hskip 1.5  true cm 3. The Pomeron in QCD

\hskip 1.5  true cm 4. The Pomeron in deep inelastic scattering

\hskip 1.5  true cm 5. The Pomeron structure

\hskip 1.5  true cm 6. (Temporary?) Conclusions

\hsize=17 true cm
\leftskip= 0 true cm
\vfill\eject
\baselineskip 18 truept
\hsize 17 true cm
\vsize 23 true cm

{\chap  1. Introduction. How many Pomerons?}
\bigskip

Recent times have witnessed a tremendous revival of interest in diffraction
and related issues. A field that had been agonizing for decades, now,
has been revitalized by the injection of the great excitement coming from the
new HERA experimental data and
from the realization that low-$x$ physics in deep inelastic
phenomena is strongly and almost directly related to the hadronic
high energy soft ({\it i.e.} low $p_T$) physics known as {\it diffraction}.
 Hundreds of papers and tens of different approaches have focused especially
on the meaning, origin and structure of the {\it Pomeron} {\it i.e.} of the
agent credited to mediate diffraction.

\medskip

This paper can only partly be considered a {\it real} review paper on
diffraction or on the Pomeron; more properly, it is the discussion and the
review of a selection of presently hot topics related to the problem of {\it
diffraction and the Pomeron} in which the present authors have been active for
a long time. As such, the perspective in which this paper is conceived reflects
mostly the view of its authors; even so, we believe that the presentation of
this material is pedagogical enough and kept to such a general level as
to be especially helpful to the reader who would like to see these fundamental
issues reduced to the essential and discussed in as simplified context as
possible. The reason why this is not a full size review paper is that,
after giving the matter a long and careful consideration, we have decided
that the huge number of papers appeared in recent times, would have made it an
impossible task that of writing a {\it bona fide}, unbiased review paper
in which the proper credit would have been given to all authors and approaches.
The state of the art is presently such that one should really plan to
write a {\it series} of review papers each focusing on a different
aspect of the problem. The field is in such a rapid swing that it would
be impossible to do a good job nor is it possible at this point to predict
whether or when the dust will have settled enough as to allow one a definite,
unemotional, unbiased and complete vision of the underlying physical problems.
All complications arise, obviously, from the fact that conventional diffraction
is, inherently and unavoidably, non perturbative and we have not yet learned to
master non-perturbative quantum chromodynamics. In the course of the paper,
as we mentioned, we will focus on several issues related to the various ways
diffraction manifests itself and we
will go deeper in some of them. While we will try to present these issues in as
wide as possible perspective, we acknowledge that we will largely ignore
the point of view of several very active physicists in the field. We apologize
from the start to all those that will feel (rightly so) that their work was not
adequately covered; as we have explained, this would not have been possible.
The reader who would like to broaden his perspectives beyond what we can do in
this paper, is urged to consult, for instance Ref. [1] where a large spectrum
of the present days ideas on diffraction was presented and discussed.

\medskip
\vfill\eject

There is much confusion about the Pomeron ($\equiv$ diffraction).
One of the most persistent is
that there are two Pomerons - a {\it soft} and a {\it hard} one. The two-fold
interpretation of a single object (phenomenon) however has a clear origin :
the conventional Pomeron studied in hadronic physics (as mentioned already,
we take this term to be equivalent to {\it diffraction}) is a {\it soft}
phenomenon, outside the range of applicability of perturbative quantum
chromodynamics, what has come to be known as the {\it hard} Pomeron is
something that can be calculated from perturbative quantum chromodynamics.
The recent small-$x$ data collected at HERA were interpreted by some
authors as a manifestation of the {\it hard} Pomeron, and thus as an argument
in favor of the existence of two Pomerons.

\medskip

In our opinion there is only one object and the diversity of its manifestations
reflects merely the diversity of the reactions in which it can occur and of the
physical and kinematical situations in which it is investigated. More
specifically, we visualize the Pomeron as a very complicated entity (which
will,
in fact, be a function of different sets of variables depending on the reaction
one is looking at) which in different dynamical situations may have different
manifestations but whose origin is always the same, diffraction,
\medskip

\hsize=16 true cm
\leftskip= 0.75 true cm
{\it "In my opinion, the Pomeron - as a leading
singularity in the $j$-plane - is unique, but it contains contributions both at
large and small distances. The relative weight of these contributions depends
on the given process (at given energies) and hence there is no universality.
These contributions mutually renormalize and without a scale it is impossible
to determine which one is more important."}\footnote{$^1$}{\ninerm
Translated from russian and quoted
freely from a private communication to one of us, L. L. J., by Lev Lipatov.}

\hsize=17 true cm
\leftskip= 0 true cm
\smallskip
The idea of a unique Pomeron is also discussed in the recent talk of T.T. Wu
[2a] :

\hsize=16 true cm
\leftskip= 0.75 true cm

{\it "It should be emphasized that there is only one Pomeron. In particular,
the
so-called soft Pomeron and the hard Pomeron are merely aspects of the same
object."}

\hsize=17 true cm
\leftskip= 0 true cm
\bigskip
\smallskip
In the present paper, we will try to clarify these points presenting them first
in the framework of the old (but still useful) methods of the analytic
$S$-matrix theory and then combining this description with the quark-parton
picture of the hadron structure and with quantum chromodynamics (QCD). We
will, thus, start from a brief introduction of the basic properties of the
theory of complex angular momenta (by now an almost forgotten language) and
then we will rephrase the same ideas of diffraction in the modern QCD language.
\vskip 1 true cm

\vfill\eject

{\chap 2. The Pomeron in the $S$-matrix theory}
\bigskip

The Pomeron was introduced in the early sixties in the framework of the
complex angular momentum theory, to describe high energy elastic and
diffractive hadron scattering\footnote{$^2$}{\ninerm In the high energy
collision of two hadrons, the distinction between diffractive and elastic
scattering is that, in the former, one of the initial hadrons has been
replaced, in the final state by a {\itf hadronic system} whose quantum numbers
(except, possibly for the spin) are the same as those of the impinging hadron.
This gives rise to {\itf single diffraction}, similarly for {\itf double
diffraction} when the same occurs for both colliding hadrons.}.
Empirically, diffraction implies non-decreasing (with energy) total and
elastic cross sections, a very rapidly decreasing (with momentum
transfer) forward peak (usually parametrized as an exponential) and an (almost)
imaginary forward elastic scattering amplitude.

Let us make a brief pragmatic introduction to the basic ideas leading to the
formalism of complex angular momenta. For further reading on this subject we
recommend the excellent book by P.D.B.Collins [2b] and several review
papers which have appeared recently [2c].

Once the scattering amplitude is expanded in partial waves, if one
assumes that the partial wave amplitude $a(j,t)$ is dominated by an
isolated, simple, moving pole located in the complex angular momentum plane $j$
at some value $\alpha(t)$

$$a(j,t)={\beta (j)\over j-\alpha(t)}, \eqno(2.1)$$

\noindent
using a well-defined mathematical prescription known as the
{\it Sommerfeld-Watson transform} [2b], one finds that the asymptotic behavior
of the elastic scattering
amplitude A(s,t) in the limit  $s\gg m^2, t\approx 0, s\rightarrow \infty $
($s$ and $t$ are the square of the total centre-of-mass energy and
four-momentum transfer in the process respectively), is given by

$$A(s,t)=\xi(\alpha)\beta(\alpha)(s/s_0)^\alpha,\eqno(2.2)$$

\noindent
where $\xi(\alpha)$, known as the {\it signature factor}, can be written as

$$\xi(\alpha)=
{1+\xi \exp^{-i\pi\alpha}\over{\sin \pi \alpha}},\ \ \ \xi=
\pm 1.\eqno(2.3)$$

\noindent
In the above Eq. (2.2), $\beta(t)$ is known as the residue function,
$\alpha\equiv\alpha(t)$ is called {\it Regge trajectory} and
$s_0$ is a scale parameter.

\medskip

Several comments are in order.

\hskip 0.5 true cm
1. The same asymptotic result (2.2) obtains if, instead of
an isolated, simple pole (2.1), the partial wave amplitude is endowed with a
{\it finite} number of isolated moving poles of which (2.1), however, is the
one with the {\it largest real part}. This will be assumed in what follows.

\hskip 0.5 true cm
2. Simple Regge pole exchange implies that the scattering
amplitude (2.2) factorizes into a "propagator" $s^{\alpha(t)}$ and two vertex
functions (Fig. 2.1); hence the residue function $\beta(t)$ may be interpreted
as the product of the two vertices $\beta(t)=g_1(t)g_2(t).$

\hskip 0.5 true cm
3. In Regge pole models the s-dependence is (asymptotically) rather simple and
well-defined, not so, unfortunately, the t-dependence. As a matter of fact,
the residue functions in Regge pole models are in general quite arbitrary
(and so will be the dependence on other variables which may arise in more
complicated reactions; such will be the case of the $Q^2$ dependence in
{\it hard diffraction} which will be considered in Section 4). Following dual
models, we will assume that the residues depend on $t$ entirely through
the trajectories $\alpha(t)$. A simple and reasonable choice is

$$\beta(\alpha)=Ae^{b\alpha(t)},\eqno(2.4)$$

\noindent
where $A$ and $b$ are free parameters.

\hskip 0.5 true cm
4. The numerical value of the scale parameter $s_0$ in Eq. (2.2) is irrelevant
since it can be absorbed in the slope parameter $b$ in Eq. (2.4) (this will not
be true any more for more complicated Regge singularities); it is reasonable to
chose  $s_0=1$ GeV$^2$ this being the usual scale of strong interactions.

\hskip 0.5 true cm
5. Notice that the denominator in the signature factor vanishes at all integer
values of $\alpha(t)$. Correspondingly, however, the numerator vanishes for
odd and, respectively, even integers according to whether we are considering
positive or, respectively, negative signatures. The recurrent blow up of
$\xi(\alpha)$ (and, consequently, of the amplitude) every time $\alpha(t)$
crosses the appropriate (even or odd) integer, is at the basis of the
statement that a Regge trajectory interpolates among families of resonances.
We can thus
conclude that a {\it positive signature trajectory} interpolates between evenly
spaced resonances that, (all other quantum numbers being equal), have {\it
even} spin (and, conversely, a negative signature trajectory interpolates among
evenly spaced resonances
of {\it odd} spin)\footnote{$^3$}{\ninerm Here we limit ourselves to
{\itf mesonic} trajectories for simplicity but the argument could be easily
extended to baryonic trajectories.}.

\bigskip
      The signature factor $\xi(\alpha)$ in Eq. (2.3) may be approximated (for
C-parity $C=+1$ of the exchanged particle) by $\xi(\alpha)=
e^{-i\pi\alpha/2}$, whenafter the expression for a single Pomeron pole exchange
scattering amplitude (2.2) becomes extremely simple

\vfill\eject

$$A(s,t)=Ae^{B(\tilde s)\alpha(t)},\eqno(2.5)$$

\noindent
with $B(\tilde s)=b+ ln(\tilde s),\ \tilde s=-i{s \over s_0}.$

\hskip 0.5 true cm
6. The trajectories $\alpha(t)$ extrapolate between the resonance ($t>0$)
and the scattering ($t<0$) regions. They are analytic functions of their
argument $t$  with  threshold singularities [3]. As it appears from the
Chew-Frautschi plot (spins {\it vs} (masses)$^2$) which follows from fits
to the scattering data, the trajectories are nearly linear within
a fair range of $t$ (Fig. 2.2),

$$ \alpha(t) = \alpha(0)+\alpha't. \eqno (2.6)$$

\noindent
$\alpha(0)$ is known as the intercept and $\alpha'$ as the slope of the
trajectory. The Pomeron is the Regge pole with the largest intercept
(arguments have been given that, asymptotically, they must behave
logarithmically to meet wide angle scaling behavior in dual models [3]).
The fact that the lowest mesonic resonances all lie on essentially the same
straightline independently of their quantum numbers (signature included), has
come to be known in the literature as {\it exchange degeneracy}.

\hskip 0.5 true cm
7. The $S$-matrix theory, and the Regge pole model in particular
deal with external particles which are on the mass shell, $p^2=m^2.$

\hskip 0.5 true cm
8. A single pole exchange ({\it i.e.} a simple moving pole in the complex
angular momentum plane) is the simplest ingredient normally used as a basic
part of the dynamics and, as we saw in Eq. (2.2), it leads to predict a power
behavior of the scattering amplitudes. More realistic models may involve more
complicated $j$-plane structures, either generated by unitarity (Regge cuts)
and/or coming from a more involved input (like a double
pole\footnote{$^4$}{\ninerm Which we will call a {\itf dipole}. Similarly, we
will briefly call {\itf multipoles}, poles of multiple order. Increasing the
order of poles increases the power of the logarithmic growth with energy of
the cross section.}). Perturbative QCD calculations [4] indicate that the
Pomeron has indeed a complicated $j$-plane structure. Phenomenology of hadronic
reactions, on the other hand, also points, independently, towards a
complicated $j$-plane structure whereby integrated hadronic cross sections grow
as $ln \, s$ or $ln^2 \, s$ (a kind of growth traditionally attributed to
cuts in the complex angular momentum plane).

\hskip 0.5 true cm
9. Multiperipheralism is a basic ingredient behind the Regge pole theory. It
implies that the dominant production mechanism is the one shown in Fig. 2.3,
in which each particle along the chain is produced peripherally, {\it i.e.}
at small momenta transfer with respect to the adjacent one. To anticipate
the forthcoming discussion of {\it hard diffraction}, we note that this
type of multiple production mechanism is quite different from the one related
to QCD evolution and typical of deep inelastic scattering (but is,
on the other hand, quite close to the one used in the so-called BFKL mechanism
to be briefly discussed below).

\hskip 0.5 true cm
10. Synthetizing : the asymptotic behavior in the (direct)  s-channel ({\it
i.e.} when $s\rightarrow +\infty$) is given by Eq. (2.2) and is determined
by the singularity in the complex $j$-plane with the rightmost real part
{\it i.e.} by the {\it Reggeon} with the largest part of Re$\,\alpha$ exchanged
in the crossed (t and/or u) channels. This is what makes the theory of complex
angular momenta so fruitful for phenomenological analyses.

\medskip

In a truly asymptotic regime, the leading trajectory (the Pomeron, in fact)
is supposed to dominate. In realistic processes, however, {\it i.e.} at
subasymptotic energies, all trajectories whose exchange is allowed by
quantum number conservation may have to be taken into account if a good
description of
the data is desired. For example, high-energy $pp\
(\bar pp)$ scattering can reasonably well be described [5] by the
exchange of the following four trajectories

$$A^{\bar pp}_{pp}(s,t)=f\pm \omega +\pom \pm O, \eqno(2.7)$$

\noindent
where the symbols on the r.h.s of (2.7) denote the $t$ channel exchanges,
which are relevant for these reactions, the reggeons $f$ and $\omega$, the
Pomeron $\pom$ and the Odderon $O$
\footnote{$^5$}{\ninerm In other reactions (or
kinematical domains), other, additional, trajectories may have to be used
such as the ones with the quantum numbers of the mesons $\rho$, $a_2$ etc...}.

Table 1 shows the $C$- parities and
intercepts of the 4 trajectories in Eq. (2.7). Trajectories with negative
parity contribute differently to particle and antiparticle induced reactions
(unfortunately, at high energies only $\bar pp$ data exist, which makes
the identification of the  Odderon, $C$-odd counterpart of the Pomeron,
not so straightforward [5]).

The Pomeron has the same (vacuum) quantum numbers as the $f$-meson; this
makes it sometimes difficult to discriminate between
these two exchanges. The intercept of the $f$-trajectory is correlated by
the spectrum of a family of resonances lying on it (see Fig. 2.2), constraining
its intercept to be around the value $0.6$. This leaves some
flexibility in the discrimination between $f$ and $\pom$ \ ($\pom$-$f$ mixing).

General unitarity arguments [6], constrain the {\it physical} Pomeron
intercept,
$\alpha_{_{\pom}} \leq 1$. What has come to be known as the
{\it bare} Pomeron intercept, however, may be greater than one, (in order to
reproduce the rise of the total cross sections with energy). Such a {\it
supercritical} Pomeron behavior has then to be properly tamed by unitarity
before the game is over.

The best way to have a clear signal of the Pomeron is to increase the reaction
energy. In fact, for the first time, at the Tevatron the total cross
section  can be described by the Pomeron only, the relative contribution
from the secondary trajectories being smaller than the experimental
error bars. The data\footnote{$^6$}{\ninerm The reader should be warned, once
again that
serious data-fitting cannot avoid fitting together integrated cross sections
{\itf and angular distributions at the same time}. The literature is crammed
with wrong statements coming from overlooking such a necessity.} favor a
moderate increase of the total cross section, its rate ranging between
$ln \, s$ and $ln^2 \,s,$ or  - in terms of a supercritical Pomeron -
$s^\epsilon$ with $\epsilon\approx 0.08$ [7]. Large values of the parameter
$\epsilon$ \ (e.g. $\epsilon \ge 0.3$, as calculated [4] from
perturbative QCD) are ruled out even by fits to the data on total cross
sections.

Recently, total cross sections for real photon-proton scattering were
measured atHERA. Fig. 2.4 shows the data on proton, antiproton and
real photon - proton
cross sections with representative fits taken from Ref. [8]. The
importance of the HERA data on total cross sections is two-fold. On the
one hand, these data reach very high energies (comparable only to the
$\bar p p$ Tevatron  energies)
providing information about the Pomeron and, in addition, they give a
direct link between hadron- and lepton- induced reactions; on the other hand,
they allow a direct probing of composite structures (the proton) by an
{\it elementary} probe (the electron).

     An important and non-trivial argument in favor of the complex angular
momentum
theory is the observed shrinkage of the diffraction cone. From its rate,
the slope of the Pomeron trajectory is calculated to be $\alpha'
\approx 0.25$ GeV$^2$. This implies that as the energy increases, more and more
particles tend to be scattered in the forward direction (or, to be more
precise, inside a cone centered around the forward direction which narrows
progressively as the energy increases). This is a prominent feature of
hadronic diffraction. At the same time, the near universality of the slope
of {\it all} Regge trajectories {\it other than the Pomeron} makes the latter
a rather peculiar trajectory and actually hints at a possibly different (or
more complicated) origin
(diffraction) or type of complex $j$-plane singularity for the Pomeron
(may be a cut originated by unitarity rather than a simple moving pole).

      Hadron diffraction, mediated by Pomeron exchange(s) may be observed
also in single and double diffraction dissociation (Fig. 2.5) (see footnote
2) as well as in a specific configuration of multiperipheral processes called
multi-Pomeron exchange (Fig. 2.6). The common feature of all these events
is that they are described by the exchange of something with the quantum
numbers
of the vacuum, whose contribution is peaked near $t=0$ and does not decrease
with energy.

While it is rather direct to verify that non-leading Reggeons come from
singularities in the complex angular momentum since
secondary Regge trajectories are made of valence
quarks (and interpolate resonances of a conventional type, see Fig. 2.2),
the Pomeron (and Odderon) trajectory is considerably more complicated
since it is supposed to be composed mainly of gluons (eventually, with some
quark admixture). In any case, one expects that,
in analogy with secondary Reggeons, observable particles (glueballs) should be
found on the Pomeron (and Odderon) trajectory for integer values of spin larger
than one. Since the parameters of the trajectory are well defined from the
scattering region, the predictions for glueball masses (and widths, in the
case of nonlinear trajectories [9]) are quite definite. The absence of clear
experimental signals of the predicted glueballs is somewhat of an
embarrassment for the theory although some evidence has been claimed recently
[10] (Fig. 2.7).

  Deep inelastic lepton-hadron scattering (DIS) at small
values of the Bjorken variable $x$ is another class of reactions where
the Pomeron manifests itself; this point will be discussed below.
\vskip 1 true cm

{\chap 3. The Pomeron in QCD}
\bigskip

Studies of the Pomeron in QCD started in 1975 with the papers of Low
[11] and Nussinov [12], who calculated the exchange of two gluons (Fig. 3.1),
which has the quantum numbers of the Pomeron and thus may be considered as
the Pomeron Born term in QCD.

As a further step, Gunion and Soper [13] introduced a model of
Pomeron-hadron coupling by making certain assumptions concerning the hadron
structure.

In parallel and nearly simultaneously, intense studies of the perturbative
Pomeron were initiated by Lipatov [4] and his collaborators. This has,
eventually, developed in what is now called the BFKL Pomeron [14] (see Sec. 4);
(recently, very exhaustive and complete series of lectures on this subject
have appeared [15]).

Some progress has been achieved in coupling the BFKL Pomeron to hadrons, but in
the absence of an infra-red cut-off, the logarithmic slope remained divergent
at $t=0.$ Attempts [16] to resolve this problem have demonstrated that the
relevant regularization procedure may violate factorization, required to hold
(see Sec. 2) in the case of a simple pole exchange.

It may be [17] that a regularization with the correct factorization
properties is provided by the vacuum structure of QCD, related to the gluon
and fermion condensates. In this way, the Pomeron couples to the quarks
like a $C=+1$ photon and, as a consequence, the essential properties of the
scattering amplitude are determined by the parameters of the condensate,
obtained by lattice calculations.

Several generalizations and phenomenological approaches to the Pomeron have
appeared trying also to link together various aspects [18].
Attempts to include non-perturbative contributions in the Pomeron
based on models for the relevant gluon propagator have been offered [19-22].
In general, in these models the rigorous properties of
Green's functions are treated approximately in QCD. The lowest order
calculations are in qualitative agreement with the data but they fail
at higher orders [23]. This is not surprising;
it has been shown [24] that the modification of the gluon
propagator by itself is not sufficient for the non-perturbative
effects to be accounted for. It is also necessary to modify the
corresponding Feynman rules in such a way as to provide self-consistency for
the new propagator. Below, we discuss briefly this point by referring for more
details to Refs. [24, 25].

Let us consider the transverse gluon propagator in the form

$$D(k^2)={c\over{(k^2-\mu^2)^2}}-{1\over{k^2-M^2}},\ \ \ \mu^2\ll M^2,
\eqno(3.1)$$

\noindent
which is perhaps the simplest realization of analyticity requirements [26, 27].
The parameters $c, \mu^2$ and $M^2$ depend on the coupling $g$ and vanish when
$g\rightarrow +0.$ In Ref. [24] they were considered as constant and their
values determined from the predictions implied by Eq. (3.1). Gluon condensate
[19 a)] and heavy quark spectroscopy [19 b)] suggest for these parameters

$$\mu\approx\ 50 MeV, \quad M\approx\ 750 MeV,\quad
{g^2c\over{6\pi}}\approx\ 0.224\ \hbox{\tenrm{GeV}}^2.\eqno(3.2)$$

\noindent
In Ref. [19 b)] it is shown that the mass spectrum of charmonium and
bottonium is successfully reproduced with the non-relativistic potential
associated with Eq. (3.1) and with the parameters as in Eq. (3.2).

The first term on the r.h.s. of Eq. (3.1) clearly enhances low frequency
modes and
will be referred to as the non-perturbative gluon propagator (clearly,
other, more complicated forms could be suggested, but the one used here is
not only a traditional one but, mostly, serves the purpose of an illustration).

Omitting technical details, the elastic scattering amplitude (for zero mass
quarks) is calculated (approximately) at order $g^6$ and, taking into account
also the secondary $f$ and $\omega$ reggeons, one can compare the result
[24] with the experimental data for total and differential pp scattering.
The only parameter left free from spectroscopy can be fixed from
the data; this leads to
$\alpha_s={g^2\over{4\pi}}\approx0.425$ (or $c\approx0.79$ GeV$^2$).
This value is, admittedly, a little too large but the model is
sufficiently unsophisticated as to make it not unrealistic.

Fig. 3.2 shows the calculated [24] $pp$ differential cross-section
in the energy region,

\noindent
$\sqrt s=19.4$ GeV, where the Pomeron contribution is
nearly constant. The calculation, at order $g^4,$ parallels the one by
Landshoff and Nachtmann [17] and the result is in
good agreement with experiment up to $|t|\leq 0.1$ GeV$^2$. The reason for the
deviation at larger $|t|$ is mainly due to the momentum space technique used
at high energy, many $t$-dependent terms have been dropped.

For the total cross section a more complete calculation yielding
a $ln \, s$ contribution from the $g^6$ diagrams [24] gives

$$\sigma_{tot}(s)=f+w+\alpha+\beta ln {s \over s_0}+{\xi \over s},
\eqno(3.3)$$

\noindent
where $f,w$ denotes the standard contribution from non-leading reggeon
exchanges defined in Eq. (2.7).

Fig. 3.3 shows the results of a fit [24] to the $\bar p p$ data. The
values of the fitted parameters are

$$\alpha=0.365\ mb,\quad \beta=4.9 95\ mb, \quad \xi=210.03\ mb(GeV)^2.$$

\noindent
The agreement with the data could probably be improved (at the price of
additional complications) and this could slightly modify
the parameters. The point, however, is not the best possible data-fitting;
this is not so relevant,
more important is the physical meaning and consistency in the
numerical values of the parameters. Thus, from Eq. (3.3) one gets
$\alpha_s=0.386$ (or $c=0.87$ GeV$^2$ and $c$ is near its upper bound, see
[24]).

Thus, the above model gives a successful
description both of quark interaction [24] and heavy quark spectroscopy [19 b)]
with all the parameters, except $\alpha_s$, fixed.

An expansion in higher multipoles (see footnote 4) has also been considered
[28]. Reasonable fits to the $pp$ and $\bar p p$ scattering data are then
obtained. The knowledge of the fitted parameters is important for the
reconstruction of the assumed perturbative series of multipole Pomeron.
\vskip 1 true cm

{\chap 4. The Pomeron in deep inelastic scattering}
\bigskip

        Consider now the traditional lepton-hadron Deep Inelastic Scattering
(DIS) (Fig. 4.1) with the usual definition of the kinematics

$$k^2=k'^2=0,\ q=k-k',\ Q^2=-q^2, \ \nu={pq \over m},
\ x={Q^2\over 2m\nu}, $$
$$\eqno(4.1)$$
$$ W^2=(p+q)^2=Q^2(1/x-1)+m^2.$$

\noindent where $\nu$ is the energy of the virtual photon and $x$ the Bjorken
variable. The relevant cross section is [29]

$${d^2\sigma\over{d\Omega
dE'}}={4\alpha^2E'^2 \over Q^4}[2W_1(\nu,Q^2)\sin^2(\theta/2)+
W_2(\nu,Q^2)\cos^2(\theta/2)].\eqno(4.2)$$

\noindent $\theta$ being the scattering angle between the directions of motion
of the initial and final leptons. The structure functions $W_1$
and $W_2$ are related to the forward virtual
Compton scattering and, in analogy to hadronic reactions, one may assume
Regge asymptotic behavior to hold for virtual Compton scattering

$$W_1(\nu,Q^2)\sim\nu\sigma_T\rightarrow\sum_i\beta^i_1(Q^2)
\nu^{\alpha_i(0)},$$
$$\eqno(4.3)$$
$$\nu W_2(\nu,Q^2)\sim (\sigma_T+\sigma_L)\rightarrow\sum_i Q^2\beta_2^i(Q^2)
\nu^{\alpha_i(0)-1}.$$

\noindent
According to Eq. (4.3), Regge asymptotics holds when $\nu$ is large. It is
compatible with Bjorken scaling if $Q^2$ is also large and the "residue"
behaves like $\beta(Q^2)\sim Q^{-2\alpha(0)}$ (a dimensional overall
constant has been dropped in (4.3)). The applicability and
limitations of the Regge pole model for virtual particles
scattering is a complicated and yet not completely resolved problem.

In the limit $Q^2\rightarrow 0$, we have real photon-hadron
scattering, which, by vector meson dominance, can be related to hadron-hadron
elastic scattering - a classical area for the application of the Regge pole
model. The structure functions are related to the virtual Compton scattering
total cross section by

$$\sigma^{\gamma^* p}={4\pi^2\alpha\ (1+4m^2x^2/Q^2)\over{Q^2(1-x)}}F_2(x,Q^2).
\eqno(4.4)$$

\noindent
As $Q^2\rightarrow 0$, the structure function should therefore vanish like
$Q^2$ in order to satisfy gauge invariance. This is a very delicate limit
and we shall come back to it.

By unitarity, the squared modulus of this process is equal to the
imaginary part of the forward virtual Compton scattering amplitude (Fig. 4.2).

While the diagram of Fig. 4.1 corresponds to the usual case of (fully
inclusive) DIS, in recent times, HERA has provided
lots of very good quality data on the semi-exclusive reaction of Fig. 4.3
{\it i.e.} $e \, + p \rightarrow e' \, p' \, X$ (where $e'$ and $p'$ are the
scattered electron and proton respectively and X is {\it a priori} any
hadronic system).

This is a much richer reaction which for large values of the total $\gamma^* p$
energy $W$ defined in Eq. (4.1) ({\it i.e.} for small values of $x$) is
dominated by
the exchange of the appropriate reggeons. If, in particular, the quasiparticle
X has the same quantum numbers of the virtual photon $\gamma^*$, then, for
large values of $W$, the reaction is dominated by Pomeron exchange. Notice
that,
according to Eq. (4.1), W will be large when $1/x$ will be large {\it i.e.} in
the domain of small $x$ which is precisely where HERA is at its best.
Aside from $t$ (as previously defined, $t = (p-p')^2$ is now
the square of the four-momentum transfer of the diffractive reaction
$\gamma^* p \rightarrow
X\ (J^{P,C}=1^{-,-})\ p'$), we must now introduce also $M_X^2$ (the squared
mass
of the hadronic system) and, rewriting $x={{Q^2}\over {2 p q}}$ $={{Q^2}
\over {W^2 + Q^2 - m^2}}$, it has become common practice to introduce also

$$x_{_{\pom}}= {{q (p-p')} \over {q\ p}}
={{M_X^2 + Q^2 -t}\over {Q^2}} x,\eqno(4.5) $$
and
$$\beta = x/x_{_{\pom}}\eqno(4.5')$$

\noindent as the appropriate variables which describe the Pomeron kinematics
within the reactions under study.

As we have discussed, in the limit of small-$x$ the corresponding energy
variable of the quasi-hadronic reaction $\gamma^* p \rightarrow Xp'$ becomes
large and, in this limit, the complex angular momentum analysis applies and
Pomeron exchange dominates; this is the main subject of this discussion.

In view of a previous remark that even the number of variables which define
{\it Pomeron exchange} may depend on the reaction one studies, it is important
to notice the following different choices : in the regime of non-negative fixed
$q^2$ ($m^2$ for hadron-hadron, or 0 for real photon-hadron scattering) and
$t$ variable, we have an (on shell) elastic scattering amplitude in the
framework of the analytic $S$-matrix theory (see Sec. 2); for $t=0$ but
variable (negative) $q^2$ we have a case of current-hadron interactions of
which the most typical example is DIS, (in this case, as recalled, we use the
positive variable, $Q^2=-q^2$); finally, for variable
(negative) $q^2$ ({\it i.e.} positive $Q^2$) {\it and} $t$ also variable, (or
$x_{\pom}$ different from $x$), we have the novel experimental situation
first encountered at HERA : exclusive (or semi-exclusive) vector meson
(virtual)
photoproduction over a wide range of $Q^2,$ {\it i.e.} a reaction in which all
three variables $W^2$ (or $s), t$ and $Q^2$ (or a different choice of the
previously defined ones) enter as independent variables in the same reaction.

It is the latter regime of Pomeron dominance which has
come to be known in the literature as {\it hard diffraction}. The terminology
arises from the fact that, the experimental investigation is performed in a
region in which enough energy is transferred to create a hadronic state $X$
(or, at FERMILAB, jets) and, yet, the final proton retains almost all of its
initial momentum {\it i.e.} is basically not deflected from its original
direction so that, to all effects, we are still in a domain dominated by
{\it soft} physics (the $\gamma^* p \rightarrow X p$ reaction) but this is
the result of a {\it hard} interaction whereby the state $X$ (or a jet) has
been emitted at large $p_T$. All this was made possible within a lepton-hadron
DIS by having access to sufficiently low $x$'s.

As it is obvious, most of the discussion that follows is confined to the
region of hard diffraction {\it i.e.} to the domain of very small $x$; thus,
here, we will totally ignore another interesting limit, $x\rightarrow 1,$
which corresponds to the transition between DIS structure functions (SF) and
elastic form factors. Near $x=1$, the SF are far from the Regge domain and
their parametrizations obey the quark counting rules. The interested reader
may consult Refs. [29] or [30] for more details.

In Fig. 4.4 we show a typical road map, quoted (with liberal modifications)
by various authors, showing different regimes in the behavior of the SF.
Here the figure is intended only as a pictorial help to guide the reader.
In fact, the definition of these regions and, mostly, the transitions and the
connections between them are still far from clear and will be only very
briefly discussed below.

A quite general question that imposes itself in the light of the above
discussion is : what is the real range of applicability of the Regge pole model
in DIS? Intriguing and controversial is the transition in $Q^2$ between
the assumed Regge behavior and the QCD evolution. Is this transition
smooth, or abrupt (like a phase transition)? Are the two regimes compatible,
{\it i.e.} does an interface between them exist?

Formally, we know of no upper limit in $Q^2$ for the Regge pole theory to be
applied in DIS (but, on the other hand it is not obvious that the usual high
energy theorems should hold at $Q^2 \neq 0$). This attitude has been taken in a
recent paper [31], where a direct fit to the data including those at the
highest $Q^2$ values by a Regge behaved SF was attempted.

To insure Regge behavior to hold in the various kinematical regions we have
so far considered, we recall that $W^2 = (q+p)^2$ should
be $ \gg 0$ (corresponding to the Mandelstam energy variable $s$ in elastic
hadron scattering being large, see Fig. 2.1). As remarked already, $W^2$ large
requires $x$ small which, together with $Q^2$, is the variable in which the
data are normally presented. For this to be possible,

i) the interaction energy ($\sqrt s$) must be large enough to ensure that $W$
can be far from the resonance region {\it i.e.}

$$W^2\gg m^2,\eqno(4.6)$$

\noindent
where $m$ is a typical hadronic mass. Since $W^2=Q^2(1/x-1)+m^2$, if we take
$x$ sufficiently small, one has

$$Q^2/m^2\gg x;\eqno(4.7)$$

\noindent
ii) The cosine of the scattering angle in the $t-$channel should be large. In
conventional DIS, $t=0$; therefore

$$|cos\theta_t|={W^2+Q^2-m^2\over{2\sqrt{Q^2m^2}}}\gg 1,\eqno(4.8)$$

\noindent
requiring

$$Q^2/m^2\gg x^2.\eqno(4.9)$$

\noindent
The first condition is the strongest if $x \neq 1$. Hence, Regge behavior is
expected in the region right and upwards from the boundary

$$Q^2\geq{x\over {1-x}}(W_{min}^2-m^2),\eqno(4.10)$$

\noindent
shown in Fig. 4.5 where the value $W_{min}= 3$ GeV may be a reasonable
choice.

In analogy with hadronic processes, one assumes that the total cross
section

$$\sigma=\sigma_L+\sigma_T=ImA(W^2,t=0,Q^2)\eqno(4.11)$$

\noindent
is Regge behaved in the domain specified above.

In $\gamma p$ scattering, by quantum number conservation, only positive
C-parity exchange is allowed, so one can write (see Table 1) :

$$A=\pom + f,\eqno(4.12)$$

\noindent
where $\pom$ and $f$ stand for Pomeron and f-meson exchange.

An immediate consequence of a purely Regge-type fit (as discussed in [31])
is that, in spite of the large flexibility inherent in Regge pole models
(especially, in the parametrization of the residue function), it seems very
hard to reconcile the mild increase with energy of the cross sections at
small $Q^2$ with the rapid increase found for $Q^2$ around $5$-$10$ GeV$^2$
(Fig. 4.6). A possible conjecture is that the latter may not have a
conventional Regge-like origin, for instance it could be attributed to
a $Q^2$ dependent slope of the Pomeron (two explicit mechanisms, one assuming
it and the other trying to provide an explanation for it, are briefly discussed
later in Section 4) or, as we will see in Section 5, this effect could also be
blamed to the QCD evolution.

Beyond the (non-perturbative) Regge  domain, one expects perturbative QCD to
be applicable.

One distinguishes two typical perturbative regimes, namely :

1. If one assumes
$$\alpha_s(Q^2_0) ln(Q^2/Q_0^2)\sim 1, \ \alpha_s(Q^2_0)\ll 1, \eqno(4.13)$$

\noindent
in the leading log($Q^2$) approximation (LLA) only the terms proportional to
\-$\alpha_s^n ln^n(Q^2/Q^2_0)$ are retained. The relevant evolution equation
($Q^2$-evolution) is named after various authors DGLAP [32]. Below we will
show how it works in the HERA domain.

2. A different perturbative regime named also after several authors BFKL [14]
leads to an ($x$-evolution) equation by assuming that,

$$\alpha_s(Q_0^2) ln(1/x)\sim 1,
\quad \hbox{with} \quad \alpha_s(Q_0^2) \ll 1\eqno(4.14).$$

\noindent
In the perturbative decomposition, only leading terms in $\alpha^n_s(Q_0^2)
ln^{n}(1/x)$ are retained. The approximate solution of this equation,
for $\alpha_s(Q^2)=\alpha_s$ fixed, leads to

$$F_2(x)\sim x^{-\Delta},\ \Delta=\alpha_{_{\pom}}(0)-1=(12/\pi)\alpha_s
ln2>0.3\eqno(4.15)$$

\noindent
which is called the {\it Lipatov Pomeron} (remember that for small
$x,\ x\sim {Q^2 \over s}$). It is precisely the relevance of this
{\it perturbative} (or {\it hard}) Pomeron in {\it soft} reactions, {\it i.e.}
when $Q^2$ is small which is presently the subject of a hot debate still quite
unsettled on which we shall focus our attention.
\medskip
Restating in different words the situation outlined above, while
the DGLAP evolution proceeds at constant $x$ and increasing
$Q^2$ (which corresponds to increasing the resolution with which the hadronic
target is probed by the photon), the BFKL equation describes the evolution
towards decreasing values of $x$ and links together {\it hard} DIS with
{\it soft} diffraction. As a consequence, the results of a calculation
depend essentially on two things, the first, over which we have little quality
control is the choice of the non-perturbative input (its form, its range
etc.) and the second, which, on the contrary we can completely control, the
procedure of evolution chosen in such a calculation.

To show an example of what we mean, consider the following nonperturbative
input SF [33]

$$F_2(x,Q^2)=A\bigg(1+\epsilon\ ln{1\over x}\bigg)\eqno(4.16)$$

\noindent
based on the expected [34] Regge pole behavior and Pomeron dominance
in the small $x$ and moderate $Q^2$ region and on the observation [35] that
according to the experimental data of NMC [36], the SF $F_2$ at $Q^2=5$ GeV$^2$
and in the range of $10^{-2}\leq x\leq10^{-1}$ increases like $ln(1/x).$
The parameter $\epsilon$ in (4.16) according to a fit [35] to the NMC data
is $\epsilon \approx 0.1$.

An explicit, approximate solution of the DGLAP equation in the form

$$F_2(x,Q^2)=B\exp(-4.74\xi)\bigg(x^{-1.44\xi}+\epsilon\ ln (1/x)
\exp(3\xi)x^{-0.72\xi}\bigg),\eqno(4.17)$$

\noindent where

$$\xi=ln{ln(Q^2/\Lambda^2)\over{ln(Q^2_0/\Lambda^2)}},\eqno(4.18)$$

\noindent
was obtained in Ref. [37], where it was shown also that the solution is
compatible with the DGLAP equation only if the power term, implying a
{\it supercritical Pomeron} is present in (4.17). To reconcile it with the {\it
dipole Pomeron} form (4.16) of Ref. [33] at $Q^2=5$ GeV$^2$,  one chooses the
scale parameter $Q^2_0$ equal to 5 GeV$^2$, at which point (4.17) reduces to
(4.16). This is the starting point of the evolution, wherefrom the behavior of
the SF at higher values of $Q^2$ is calculated in a parameter-free form. The
results are compared with the HERA data [38, 39] in Fig. 4.7 .

The virtue of this exercise is its simple analytic form. Calculations
and fits with more involved inputs and/or more sophisticated or complete forms
of the evolution can be found e.g. in Refs. [40, 41] (see also Refs. [15, 18]).

In 1993 two experimental groups at HERA  - H1 and ZEUS - measured the proton
structure function in the hitherto unexplored region
$10^{-4}\leq x \leq 10^{-2}$ and 5 GeV$^2$ $\leq Q^2\leq 10^5$  GeV$^2$.
The results were first reported at the Moriond meeting in March 1993 [38].
What became known as the {\it HERA effect} is the unexpectedly rapid increase
of the proton structure function which occurs with decreasing $x$. The effect
becomes more pronounced as $Q^2$ increases and in a wide range of $Q^2$ it can
be parametrized, for instance (but certainly not only) as $\sim x^{-\delta}$
with $\delta \ge 0.3$. By virtue of the relation $x\sim Q^2/s$ (which follows
from (4.1) at small $x$), this implies that a possible parametrization of the
HERA data is obtained with

$$F_2(x)=x^{-\delta},\ \delta \ge 0.3.\eqno(4.19)$$

\noindent
For what said above, if one assumes Regge behavior and Pomeron dominance,
$\delta$ can be related to the supercritical Pomeron intercept
$\alpha_{\pom}(0)=1+\delta.$ The parameter $\delta$ can then be plotted by
fitting the small $ (<10^{-2}) \ x$ data for the various available bins in
$Q^2$. We will comment soon on this procedure but let us for the moment
proceed as if no objections could be raised to it. In this case, the result
one obtains [42] is shown in Fig. 4.8. This result is sufficiently startling
that one may comment on it before even criticizing the procedure : two
important and possibly highly non-trivial phenomena are evident from this
figure. One is the large (0.3 - 0.4) value of the "Pomeron intercept" at large
$Q^2,\  (Q^2>10$ GeV$^2$), (a value tantalizingly close to the value (4.15)
of the {\it Lipatov Pomeron)} and the other is the very rapid variation of
$\delta$
around $Q^2=5$ GeV$^2$ . These phenomena have raised several questions that
have become central in the modern studies of diffraction in the context of
QCD such as,

1) Is the large $Q^2$ behavior seen at HERA a direct manifestation of the
{\it Lipatov Pomeron}? The need for an experimental verification of the results
of
the perturbatively QCD-calculated Pomeron and the agreement with the predicted
value of the perturbative Pomeron intercept (see Eq. (4.15)) makes
this interpretation attractive. On the other hand, other explanations may
be possible (see below) and general theoretical arguments have been given
against the {\it Lipatov Pomeron} regime in connection with the HERA data [43].

2) Is the rapid change of $\delta$ in $Q^2$ due to a phase transition? [42]
and/or is it suggesting that there exist two different Pomerons - a {\it soft}
one (active at small $Q^2$) and a {\it hard} one (active at large $Q^2$)? (this
is, really, just another way of restating the previous question).

3) What is the smooth transition between real ($Q^2=0$) and virtual
($Q^2 \gg 0$) photon scattering?

In what follows we shall address these questions but the reader should be
warned that we are far from a unique answer nor do we have a unified model
valid in all kinematical regions; for the time being, only fragmentary
understanding of these phenomena is slowly emerging.

Before proceeding further, we have to come back to the cautionary remarks
made previously when commenting Fig. 4.8. It is quite clear that treating
the data the way it was done to obtain Fig. 4.8 most probably oversimplifies
the complexity of the problem; this procedure should be critically revised.

The crucial point is that as $Q^2$ tends to zero, the SF should vanish like
$Q^2$ in order to satisfy gauge invariance. This problem has been treated in
a number of papers [8c, 42 ,44]. In particular, let us recall that an early
attempt was made to satify this condition by using a modified dipole
Pomeron [45]

$$F_2(x,Q^2)=A\bigg[1+\epsilon\ ln\bigg(Q^2({1\over x}-1)+M^2\bigg)\bigg]
 ln\bigg(1+{Q^2\over{Q^2+a^2}}\bigg). \eqno(4.20)$$

\noindent
Fig. 4.9  shows a representative fit to the data at low and intermediate values
of $Q^2.$ This model, however, similar to that of Donnachie and Landshoff
[44 b)] based on a {\it supercritical Pomeron}, does not reproduce the rapid
rise of the SF at large $Q^2,$ that may be attributed to the evolution and
which was, on the contrary, predicted by other models [41] (see, for instance
Fig. 4.10).

A generalization to the above attitude has been taken by Capella et al. [46],
who choose a Regge pole model for small $Q^2$ endowing the effective Pomeron
intercept with a $Q^2$ dependence of the form

$$F_2(x,Q^2)\sim x^{-\Delta(Q^2)}, \ \Delta(Q^2)=\Delta_0\bigg(1+
{2Q^2\over{Q^2+d}}\bigg),\eqno(4.21)$$

The resulting structure function, assumed to be valid at small and moderate
$Q^2$

\noindent $(0\leq Q^2\leq 5$ GeV$^2$) is

$$F_2(x,Q^2)=Ax^{-\Delta(Q^2)}(1-x)^{n(Q^2)+4}\bigg({Q^2\over{Q^2+a}}\bigg)^
{1+\Delta(Q^2)}+$$
$$Bx^{1-\alpha_R}(1-x)^{n(Q^2)}\bigg({Q^2\over{Q^2+b}}\bigg)^{\alpha_R}.
\eqno(4.22)$$

\noindent where

$$ n(Q^2)={3\over2}\left(1+{Q^2\over{Q^2+c}}\right)$$

\noindent to obtain at $Q^2$=$0$, the same power $(1-x)^{1.5}$ as in the Dual
parton model, $\alpha_R$ being the secondary reggeon intercept.

\noindent
{}From this input the structure function has been calculated for higher values
of $Q^2$ by means of the QCD evolution. The resulting curves are shown in
Fig. 4.11.

A further refinement is found in Ref. [47], where it is shown that, actually,
the data at not too large $x$, ({\it i.e.} for $x \le 10^{-2}$) can be
reconciled with a generalized dipole (or tripole)\footnote{$^7$}{\ninerm Which
is the same as saying an increase (at $Q^2=0$) compatible with either
$ln \,(1/x)$ (dipole form) or with $ln^2 \, (1/x)$ (tripole form), see
footnote 4.} Pomeron {\it i.e.} which behaves in such a way as not to violate
the Froissart-Martin
bound [6] while reducing to a power behavior of the Donnachie Landshoff (not
Lipatov) type for large $Q^2$. Specifically, in [47] one considers (again this
should just be taken as an example) either

$$ F_2(x,Q^2) = A_{\pom} \bigg[{\tilde x^{\epsilon(Q^2)}-
(1 + \epsilon(Q^2) ln(\tilde x)) \over {{1\over 2}\epsilon^2(Q^2)}}\bigg]
ln\bigg( 1+ {Q^2 \over {Q^2+a^2_{\pom}}}\bigg) \eqno (4.23)$$

\noindent
or, alternatively

$$ F_2(x,Q^2) = A_{\pom} \bigg[{\tilde x^{\epsilon(Q^2)}- 1
\over {\epsilon(Q^2)}}\bigg]
ln \bigg( 1+ {Q^2 \over {Q^2+a^2_{\pom}}}\bigg) \eqno (4.24)$$

\noindent
where $\tilde x = {W^2 \over s_0}$, with the hadronic scale taken as $s_0=1$
GeV$^2$. These forms reduce to the expected behavior if $\epsilon (Q^2)$
vanishes as $Q^2 \sim 0$ because $ln(\tilde x) \simeq ln ({1 \over x})$ when
$W^2 \gg Q^2$. Choosing for the slope $\epsilon (Q^2)$

$$ \epsilon(Q^2)= {\lambda \over ln2} ln \bigg(1 + {Q^2 \over Q^2 +
b^2}\bigg), \eqno (4.25)$$

\noindent Eq. (4.23) leads to a $ln^2 ({1 \over x})$ behavior in agreement with
the Froissart-Martin bound and Eq. (4.24) to a $ln ({1 \over x})$. In these
equations the parameters $A_{\pom}$ and $a^2_{\pom}$ are fixed by the
requirement that the total photoproduction cross section is correctly
reproduced. With the specific choice (4.25) made for $\epsilon (Q^2)$, there
are just two adjustable parameters $\lambda$ and $b^2$. Fitting the small-$x$
data of HERA (specifically up to $x \le 5.10^{-3}$), the result is shown
for the case of Eq. (4.23) in Fig. 4.12 with the following values of the
parameters $\ A_{\pom}$ = 5.72 10$^{-3}$, $\ a_{\pom}\ $ = 1.12

\noindent (obtained from a
fit to the total photoproduction cross section), $\lambda$=0.254, and
$b^2$=0.198 with a $\chi^2(/dof)$(/58 HERA data) of about 1.2 ( obtained from
HERA data). With the use of

\vfill\eject
\noindent Eq. (4.24) the results are similar. Two interesting
conclusions are in order. First, recall that the HERA data with
$x \ge 5.10^{-3}$ are not included in the fit; in spite of this, it is only for
very high $Q^2$ that the curve deviates considerably from the data. Second, the
asymptotic value of $\epsilon$ as $Q^2$ grows to $\approx 2000$ GeV$^2$ is
roughly = 0.3 {\it i.e.} it reaches the lower limit of what are considered the
range of values appropriate for the {\it Hard Pomeron} (the value of the
{\it soft Pomeron} \`a la Donnachie and Landshoff, 0.08 being reached for $Q^2$
between 1 and 5 GeV$^2$). If a factor $(1-x)^{\beta(Q^2)}$ correcting
for $x$ not so small is inserted in Eq. (4.23) where one takes

$$\beta (Q^2) = \beta_0 + \beta_1 \tau \qquad {\hbox{\tenrm{with}}} \qquad
\tau = ln \bigg({ln ({(Q^2+Q^2_0) \over \Lambda^2}) \over ln({Q^2_0 \over
\Lambda^2})}\bigg), \eqno (4.26) $$

\noindent the values of the various parameters are now $A_{\pom}$ = 5.72
10$^{-3}$, $a_{\pom}$ = 1.12, $\lambda$=0.256, $b^2$=0.21, $\beta_0$=7.0 and
$\beta_1$=5.6 with a $\chi^2(/dof)$(/67 HERA data) of about 1.55.  Notice that,
as expected, only the parameters involved in $\beta(Q^2)$ are sensitive to
including larger $x$-values in the fit. Fig. 4.13a shows the equivalent of
Fig. 4.12 {\it i.e.} the variation in $Q^2$ for the various bins in $x$)
whereas Fig. 4.13b shows the converse {\it i.e} the variation in $Q^2$ for the
various bins in $x$. We conclude that the large disagreement in Fig. 4.12 was
due to the lack of an appropriate treatment of the not so small $x$ data.

The approaches of Refs. [42, 45-47], although different in many and also
important details, are similar in that a Regge-behaved SF was used in the
region of small and intermediate values of $Q^2$. In addition, they all go
towards the unifying spirit of saying that a unique form for the Pomeron is
responsible of both its {\it soft} and its {\it hard} aspects.

Probably, however, the real message of all this discussion is that, in spite
of $x$ being so close to zero, the HERA data are not more asymptotic than
the corresponding hadronic data and the latter are well known to be capable
of being fitted with either a logarithmically increasing or a power-like
cross section. As experience with hadronic physics has taught us [48], it is
only when angular distributions are also used in the fit that one can really
discriminate between a power-like and a logarithmic behavior. Possibly, the
same will be the case for the DIS data; we can only hope that angular
distributions from ZEUS and H1 will be available soon.

The $Q^2$ dependence of the Pomeron intercept conjectured in Refs. [46, 47]
raises an important question, namely that of factorization. We recall (see
Sec. 2) that, due to factorization, the scattering amplitude for the exchange
of a Regge pole is the product of the vertices and of the {\it propagator}
$s^\alpha$. The trajectory $\alpha$ in this case should not depend on the
masses (or virtualities) of external legs. A (presumably) mild and smooth
$Q^2$-dependence is expected from t-channel unitarization {\it i.e.} from
rescattering corrections. In principle this is calculable although technically
very complicated; the calculated multi-Pomeron exchanges depend both on the
propagator and on the vertices, involving $Q^2$ dependence. Realistic and
quantitative estimates of the expected effect as well as its experimental
verification have not yet been carried out. In addition, s-channel
unitarization (which is prohibitively complicated to enforce), is probably
even harder to estimate. Empirically, we can only interpret the {\it effective}
$Q^2$-dependent intercept of the Pomeron as an effective way of taking into
account unitarity {\it i.e.} as the collective effect of rescattering
corrections. In this sense, it could be argued that the Pomeron is an
atypically complicated angular momentum plane singularity (this, however,
has been long recognized, as we have already mentioned earlier).

An original explanation of the apparent $Q^2$-dependence of the Pomeron
intercept has been suggested in Ref. [49] where it is shown that the origin of
this effect (and the resulting non-factorizability of the Pomeron) may be that
(contrary to the common belief) the singlet SF, $F_2^S$, has two different
components - one coming from the sea quarks and the other one due to the
gluons via the $g \mapsto q\bar q$ process. As a consequence, the singlet SF
may be written in the form

$$ F_2^S(x,Q^2)\sim\alpha(Q^2)x^{-\delta_\Sigma}+\beta(Q^2)x^{-\delta_G}
\eqno(4.27).$$

\noindent
{}From a fit to the data (see Fig. 4.14) the values of the parameters were
found [49] to be $\delta_\Sigma=0.2$ and $\delta_G=0.4.$ The sum of the two
terms may be simulated by $x^{-\delta(Q^2)},$ where $\delta(Q^2)$ varies
from 0.2 to 0.4 around $Q^2=5$ GeV$^2$. (A similar point of view, although in a
different context, has been taken recently by Martin {\it et al.} [50])

\vskip 1 true cm

{\chap 5. The Pomeron structure}
\bigskip

The studies of the Pomeron structure or, alternatively, of {\it hard
diffraction} have become among the most topical recent subjects. Physically,
the motivation for these studies relies on the idea [51] that the Pomeron is a
quasi-particle with its own structure, that might be resolved in deep
inelastic photon-Pomeron scattering  in which it is assumed that
$X$ is a $1^{--}$ system (see the discussion preceding Eq. (4.5)).

Formally, the subject is nothing but diffraction dissociation, similar to the
case of hadronic reactions (Fig. 2.5), a new degree of freedom being the photon
virtuality. For this reason, the appearance of a rapidity gap in virtual
photon-proton scattering, observed at HERA [52] (Fig. 5.1) and at the TEVATRON
[53], by itself is not surprising but it has been an important signature in
order to make sure that we are indeed in the presence of diffraction [54]. What
is really intriguing, is the possibility that perturbative QCD calculations
may eventually be relevant to the understanding of the {\it soft} Pomeron.
Before presenting a particular model for the Pomeron structure and its
observable consequences, let us briefly discuss the more general problem of the
transition from the small-$Q^2$ Regge behavior to the large-$Q^2$ QCD
evolution. Since $W^2 \simeq Q^2({1 \over x}-1),$ one can write the rapidity
intervals additively as shown in Fig. 5.2. The lower part of the diagram
corresponds to a multiperiphal mechanism of particle production, which by
unitarity is related to Regge pole (in particular, Pomeron) exchanges, while
its upper part corresponds to the Reggeon's (e.g. Pomeron's) inelastic vertex,
described by the QCD evolution equation. The transition between the two regimes
may be abrupt, similar to a phase transition. The dependence of the {\it
effective Pomeron intercept} on $Q^2,$ as extracted from a fit of the structure
function  $\sim x^{-\alpha(Q^2)},$  is instructive in that it shows the
{\it phase transition} between the low $Q^2,$ corresponding to a Pomeron
exchange and the large $Q^2$ behavior coming from the QCD evolution in the
upper vertex of Fig. 5.2. It is evident from this picture that there is no
{\it hard} Pomeron - the large $Q^2$ behavior comes from the evolution of the
upper (inelastic) Pomeron vertex, while the intercept of the Pomeron itself is
almost $Q^2$-independent (a mild $Q^2$-dependence is the result of the
rescattering or unitarity corrections). The crucial question is {\it what (if
any) is the memory of the} Pomeron {\it in the QCD evolution of its vertex}?

The previous discussion raises qualitatively important issues and it is,
therefore, necessary to provide evidence that our conjecture can be used as a
working hypothesis. Several attempts along these lines have been carried out
[1, 55, 56] of which we are just going to illustrate one [55]. Once again, it
is not so much the details of the example which are relevant, but the
underlying
philosophy and the light that the specific example sheds on basic questions
which matters.

Let us recall that the differential cross section of a diffractive deep
inelastic process is

$${d\sigma(ep\rightarrow epX)\over {d\zeta dtdxdQ^2}}=
{4\pi\alpha^2\over{xQ^4}}(1-y+
{y^2\over {2(1+R)}})F_2^{diffr.}(x,Q^2;\zeta,t),\eqno(5.1)$$

\noindent
where $y$ is ${{E-E'} \over E}$, $\zeta=1-x_F$, where $\ x_F=p'_z/p_z$ and
$t=(p-p')^2.$ Note that at HERA, $(k+p)^2=\sqrt s=296$ GeV and the
pseudorapidity $\eta$ of the smallest detector angle is
$\eta=4.5,\ \Theta=1.5.$ The cut in $\eta,\
\eta_{max}<1.5,$ distinguishes events with a large rapidity gap and is
equivalent to $\zeta_{max}\leq0.06.$ Due to acceptance cuts, the maximum value
of $\zeta$ is $\zeta_0=2.0\times 10^{-2}$ for ZEUS [52b]. Factorization

$$F_2^{diffr.}\rightarrow F_{{\pom} /p}(\zeta,t)
G_{q/{\pom}}(x/\zeta,Q^2)\eqno(5.2)$$

\noindent
of the structure function, where $F_{{\pom}/p}$ is the so-called {\it Pomeron
flux} (see below) and $G_{q/{\pom}}$ is the Pomeron structure function, is an
important assumption, adopted in most of the calculations of DIS.

\vfill\eject

The Pomeron flux

$$F_{{\pom}/p}={d\sigma^{diffr.}\over{d\zeta dt}}{1\over
{\sigma_{tot}({\pom}p)}}\eqno(5.3)$$

\noindent
is defined as the diffraction dissociation cross section divided by an abstract
quantity called the Pomeron-proton total cross section; the latter is assumed
to have a behavior typical of high energy hadron scattering, for instance [55]

$$F_{P/p}(\zeta,t)=[\exp(B\alpha(t))\zeta^{-\alpha(t)}]^2\zeta.\eqno(5.4)$$

\noindent
Practically, the Pomeron flux can never be completely isolated from non-leading
(secondary Reggeon) contributions.

The key point in these and similar calculations is the choice of the Pomeron
structure function. A priori, it is not known whether the Pomeron is made of
quarks or gluons (most likely - of both). One simply makes a guess,
calculates the observable consequences and then compares the results of the
calculations with the data.

In Ref. [55] the following simple choice was made for the gluon distribution
function, corresponding to a Pomeron made of a few gluons

$$zG(z,Q^2)=a(Q^2)(1-z)^{b(Q^2)}.\eqno(5.5)$$

\noindent
The parameters

$$b(Q^2)\simeq 1+4\xi/3,\ a(Q^2)\simeq\exp{\xi/3},\eqno(5.6)$$

\noindent
are calculated [55] from the evolution equation and normalized to $a(Q_0^2)=2,
\ b(Q^2_0)=1,$ where $\xi$ is the usual QCD evolution variable (4.18).

The ratio of the diffractive to all deep inelastic events

$$r={L(x,Q^2)\over{F_2(x,Q^2)}}$$

\noindent
was then calculated and compared with the HERA data. The result is shown in
Fig. 5.3.

A resume of these and similar calculations is that given the large flexibility
of the models, coming from various assumptions such as factorization, Pomeron
pole dominance etc. as well as the large experimental uncertainties, it is
difficult to make a definite discrimination between the existing models of
the Pomeron structure [56]. Nevertheless, it is very challenging to try to
apply perturbative QCD calculations in studying the structure of an object
that is essentially non-perturbative in nature.

A direct signal of the Pomeron structure could be manifest in jets with large
transverse energy produced by a hard photon-Pomeron scattering (upper part of
the diagram, Fig. 5.2).  Experimental searches for such signals are under
way both at HERA  and the Fermilab Tevatron and are included in various
proposals for the LHC.

\vskip 1 true cm
{\chap 6. (Temporary) conclusions}
\bigskip

  As implied in the title of this Section, the conclusions cannot being other
than temporary. Much better data, more conclusive analyses and, especially,
better theoretical understandings are needed before firm conclusions can be
reached.

Presently, there are two different competing approaches to the Pomeron or
high energy diffraction.

The first assumes that there are two Pomerons, one called {\it soft} the
other one {\it hard}. In this way one is able to reconcile the formalism
and the phenomenology of conventional high energy hadronic physics (see
Section 1), with an object calculated in  perturbative QCD, the {\it Lipatov
Pomeron} which might be tempting to relate to the large-$Q^2$ HERA physics.
Needless to say, the introduction  of a two-component Pomeron increases the
degrees of freedom  thus making possible excellent fits [8 c)] to the data.

The other alternative is that the Pomeron is a single (but far from simple!)
and unique object [2a, 46, 47]. The large-$Q^2$ domain (e.g. the recent HERA
data),
involving particles with large virtualities could be outside the conventional
Regge domain in which case they should be attributed to other mechanism of
which we have offered a variety ranging from $Q^2$ {\it effective} dependence
of the slope or to the parton evolution.

It seems to us that before resorting to introducing more objects to describe
the same basic physics (diffraction), one should fully examine all other
alternatives.

\noindent Differently stated :

a) It is neither economic nor aesthetic to violate the law of Occam ({\it
Entia non sunt multiplicanda praeter necessitatem}\footnote{$^8$}{\ninerm
Additional entities should not be introduced without necessity.}); before
doing so all other possibilities should be fully explored.

b) It is not clear how the two objects could interpolate; the simplest (and
most obvious) one could be to assume that the intercept is a function of $Q^2$.
This, however, leads to problems with the basic property of factorization of a
simple pole (Sec. 2). Factorization is not necessarily valid in the case of
more complicated $j$-plane singularities (like the Pomeron is very likely going
to be), but the $Q^2$ dependence has to be rather drastic to reconcile this
viewpoint with the data and anyway it is very unlikely that it is ever going
to be calculated from $t$-channel unitarity (or, even less from $s$-channel
unitarity).

c) Large-$Q^2$ may be outside the Regge domain and the power term
$x^{-\lambda}$ may have a different origin from that at low $Q^2$. The small
and large-$Q^2$ variations of the structure functions could have an entirely
different origin so that the relevant formalism may altogether be different :
complex angular momentum theory (non-perturbative) in the first case and
perturbative QCD (with evolution) in the second.

d) The uniqueness of the object called Pomeron is made stronger i) by the
recent analysis by Goulianos [57 d)] of the CDF [54 b)] and DO [54 c)]
results suggesting that, in the hard double diffraction dissociation processes,
the fraction of rapidity gap dijet events to all dijets events with equal
kinematics is the same within errors as the rate expected for soft double
diffraction dissociation in which no jets are present and ii) by the analysis
from H1 [53] of the deep inelastic diffraction results at HERA where it
appears that the same Pomeron is  involved in hard and soft collisions.

Hopefully, a transitory region will be found where the (nonperturbative)
{\it soft} Pomeron is probed by perturbative hard processes. From the practical
point of view, the main problem will remain {\it factorization}; it will be
very difficult to find where it occurs the transition from the controllable
perturbative calculations to a (model-dependent) {\it soft} Pomeron; among
other things, this will make rather ambiguous the signals of a possible
{\it Pomeron structure}.

Concerning the QCD picture of the Pomeron itself as a moving $j$-plane
singularity(ies?), this appears essentially non-perturbative and there is no
universally established consensus or generally accepted solution for the time
being. The main difficulty remains confinement and a unified treatment of the
quark interaction (scattering) and hadron structure (spectroscopy) could be
very useful. We expect that these topics will be the centre of attention in the
near future.
\bigskip
\vfill\eject

\centerline{\bf References}

\item{[1]} Proceedings of the VIth Blois International Conference on Elastic
and Diffractive Scattering (Blois, 20-24 June 1995) Ed. M. Haguenauer.
\medskip

\item{[2]} a) T. T. Wu, {\it Behavior of total and elastic cross-sections at
very high energies}, CERN-TH.7437/94;

\item{} b) P. D. B. Collins, {\it An introduction to Regge theory and high
energy physics, Cambridge University Press}, Cambridge, (1977);

\item{} c) E. Predazzi, {\it Perspectives in High Energy Physics}, Lectures
delivered at the III$^{th}$ G. Wataghin School in Phenomenology, Campinas,
July 1994;

\item{} G. Matthiae, Rep. Prog. Phys. {\bf 57} (1994) 743;

\item{} M. Bertini and M. Giffon, {\it Elastic Scattering of Hadrons at High
Energy}, Physics of Particles and Nuclei, Vol. {\bf 26} (1995) 12.
\medskip

\item{[3]} L. L. Jenkovszky, Fortsch. Phys. {\bf 34} (1986) 702.
\medskip

\item{[4]} L. N. Lipatov, Sov. Phys. {\bf JETP 63} (1986) 904.
\medskip

\item{[5]} R. J. M. Covolan, P. Desgrolard, M. Giffon, L. L. Jenkovszky, and
E. Predazzi, Z. Phys. {\bf C58} (1993) 109.
\medskip

\item{[6]} M. Froissart, Phys. Rev. {\bf 123} (1961) 1053;

\item{} A. Martin, Nuovo Cimento {\bf 42} (1966) 930; {\bf 44} (1966) 1219.
\medskip

\item{[7]} a) C. Bourrely, J. Soffer and T. T. Wu, Phys. Rev. {\bf D19} (1979)
3249; Nucl. Phys. {\bf B247} (1984) 15;

\item{} b) A. Donnachie and P. V. Landshoff, Nucl. Phys. {\bf B231} (1984) 189;
Nucl. Phys. {\bf B244} (1984) 322; Nucl. Phys. {\bf B267} (1986) 690.
\medskip

\item{[8]} a) P. Desgrolard, M. Giffon, A. Lengyel and E. Martinov, N.C Vol.
{\ 107A} (1994) 637;

\item{} b) A. Donnachie and P. V. Landshoff, Phys. Lett. {\bf B296} (1992) 227;

\item{} c) M. Bertini, M. Giffon, L. L. Jenkovszky,
{\it Small-$x$ Analysis of the proton Structure Function}, Talk given at
{\it The Heart of the Matter}, Blois, June 1995;

\item{} M. Bertini and M. Giffon, {\it A Two-component Model for the Pomeron in
Deep-Inelastic Scattering}, to appear in Int. Journ. of Phys.
\medskip

\item{[9]} P. Desgrolard, L. L. Jenkovszky and A. Lengyel, {\it Where are the
glueballs?}, in Strong Interaction at Large Distances, Hadronic Press, 1995,
Ed. L.L. Jenkovszky.
\medskip

\item{[10]} WA91 Collaboration, S. Abatzis {\it et al.}, Phys. Lett.
{\bf B324} (1994) 509.
\medskip

\item{[11]} F. E. Low, Phys. Rev. {\bf D12} (1975) 163.
\medskip

\item{[12]} S. Nussinov, Phys. Rev. Lett. {\bf 34} (1976) 1286.
\medskip

\item{[13]} J. F. Gunion and D. Soper, Phys. Rev. {\bf D15} (1977) 2617.
\medskip

\item{[14]} E. A. Kuraev, L. N. Lipatov and V. S. Fadin, Zh. Eksp. Teor. Fiz.
{\bf 72} (1977) 377 (Sov. Phys. {\bf JETP 45} (1977) 199);

\item{} Ya. Ya. Balitsky and L. N. Lipatov, Yad. Fiz. {\bf 28} (1978) 1597,
(Sov. J. Nucl. Phys. {\bf 28} (1978) 822); Sov. Phys. {\bf JETP 63} 1986, 904.
\medskip

\item{[15]} V. Del Duca, {\it An Introduction to the Perturbative QCD Pomeron
and to Jet Physics at large Rapidities}, DESY preprint 95-023;

\item{} E. Levin, {\it The Pomeron : Yesterday, Today and Tomorrow}, Lectures
given at the III$^{th}$ G. Wataghin School in Phenomenologie, Campinas, July
1994, CBPF-NF-010/95.
\medskip

\item{[16]} D. G. Richards, Nucl. Phys. {\bf B258} (1985) 267.
\medskip

\item {[17]} P. V. Landshoff and O. Nachtmann, Z. Phys. {\bf C35} (1987), 211.
\medskip

\item{[18]} See, for instance, N. N. Nikolaev and G. B. Zakharov, Z. Phys.
{\bf C53} (1992) 231; see also N. N. Nikolaev, M. Genovese and G. B.
Zakharov {\it Diffractive DIS from the generalized BFKL Pomeron} DFTT 42/94
October 1994, to be published;

\item{} N. N. Nikolaev, invited talk at the VIth International Conference on
Elastic and Diffractive Scattering, (20-24 June, 1995) Blois (France) to be
published in the Proceedings (Editions Fronti\`eres) Ed. M. Haguenauer.
\medskip

\item{[19]} a) Z. E. Chikovani, L. L. Jenkovszky and F. Paccanoni,
Yad. Fiz. {\bf 53} (1991) 526, (Sov.J.Nucl.Phys. 53 (1991) 329);

\item{} b) Z. E. Chikovani, L. L. Jenkovszky and F. Paccanoni, Mod. Phys. Lett.
{\bf A 6} (1991) 1409.

\item{} c) A. V. Kholodkov {\it et al.}, J. Phys. {\bf G18} (1992) 985.
\medskip

\item{[20]} J. R. Cudell and D. A. Ross, Nucl. Phys. {\bf B359} (1991) 247.
\medskip

\item{[21]} a) F. Halzen, G. Krein and A. A. Natale, Phys. Rev. {\bf D47}
(1993)
 295;

\item{} b) M. B. Gay Ducati, F. Halzen and A. A. Natale, Phys.Rev. {\bf D48}
(1993) 2324.
\medskip

\item{[22]} G. Preparata and P. G. Ratcliffe, Phys. Lett. {\bf B345} (1995)
272.
\medskip

\item{[23]} J. R. Cudell and B. U. Nguyen, Nucl. Phys. {\bf B420} (1994) 669.
\medskip

\item{[24]} L. L. Jenkovszky, A. V. Kotikov and F. Paccanoni, Z. Phys.
{\bf C63} (1994) 131.
\medskip

\item{[25]} F. Paccanoni, {\it Phenomenological consequences of the dipole
gluon
propagator}, Proc. of the Workshop on Elastic and Diffractive Scattering, Kiev,
 Sept. 1992, Eds. L. L. Jenkovszky and E. Martynov, p.39; {\it Quarks
Interactions and Analyticity}, Workshop on DIQUARKS II, Torino, Nov. 1992,
World Scientific, Singapore, Eds. M. Anselmino and E. Predazzi, p.225.
\medskip

\item{[26]} K. Nishijima, Prog. Theor. Phys. {\bf 74} (1985) 889; {\bf 77}
(1987) 1035.
\medskip

\item{[27]} R. Oehme and W. Zimmermann, Phys. Rev. {\bf D21} (1980) 475;
{\bf D21} (1980) 1661;

\item{} R. Oehme, Phys. Lett. {\bf B195} (1987) 60; {\bf B232} (1989) 498.
\medskip

\item{[28]} P. Desgrolard, M. Giffon and L. L. Jenkovszky, Yad. Fiz.
{\bf 56} (1993) 226.
\medskip

\item{[29]} E. Leader and E. Predazzi, {\it An Introduction to Gauge Theories
and "New Physics"}, Cambridge University Press, Cambridge, 1982. See also the
second edition {\it An Introduction to Gauge Theories and Modern Particle
Physics} (scheduled to appear before the end of 1995 always by Cambridge
Press).
\medskip

\item{[30]} S. J. Brodsky and G. R. Farrar, Phys.Rev. {\bf D11} (1975) 1309;

\item{} S. J. Brodsky and G. L. Lepage, Phys. Rev. {\bf D22} (1980) 238;

\item{} D. Sivers, Ann. Rev. Part. Sci. {\bf 32} (1982) 149.
\medskip

\item{[31]} L. L. Jenkovszky, E. Martynov and F. Paccanoni, {\it Regge
behavior of the nucleon structure functions}, Padova preprint
DFPD-TH-95-21, 1995.
\medskip

\item{[32]} Yu. L. Dokshitzer, Sov. Phys. {\bf JETP 46} (1997) 641;
\medskip

\item{} L.V. Gribov, E. M. Levin and M. G. Ryskin, Phys. Rep. {\bf 100} (1983)
1;

\item{} L. N. Lipatov, in {\it Perturbative Quantum Chromodynamics}, Ed. A. H.
Mueller, World Scientific, Singapore, 1989, p.411;

\item{} G. Altarelli and G. Parisi, Nucl. Phys. {\bf B126} (1977) 298;
\medskip

\item{[33]} L. L. Jenkovszky, A. V. Kotikov and F. Paccanoni, Phys. Lett.
{\bf B314} (1993) 421.
\medskip

\item{[34]} L. L. Jenkovszky, F. Paccanoni and E. Predazzi, Nucl.Phys. {\bf B}
(Proc. Suppl.) {\bf 25B} (1992) 80.
\medskip

\item{[35]} P. Desgrolard, M. Giffon, L. L. Jenkovszky, A. I. Lengyel and
E. Predazzi, Phys. Lett. {\bf B309} (1993) 191.
\medskip

\item{[36]} NMC Collaboration, P. Amaudruz {\it et al.}, Phys. Lett. {\bf B295}
(1992) 159.
\medskip

\item{[37]} L. L. Jenkovszky, A. V. Kotikov and F. Paccanoni, Yad.Fiz.
{\bf 55} (1992) 2205.
\medskip

\item{[38]} H1 Collaboration, C. Vallie, {\it First measurement of the proton
$F_{2}$ Structure Function at HERA}, in {\it QCD and High Energy Hadronic
Interactions}, Moriond-93, Ed. J. Tr\^an Thanh V\^an, Editions Fronti\`eres.
\medskip

\item{[39]} H1 Collaboration, I. Abt {\it et al.}, Nucl. Phys. {\bf B407}
(1993) 515;

\item{} H1 Collaboration, T. Ahmed {\it et al.}, Nucl. Phys. {\bf B439} (1995)
471;

\item{} ZEUS Collaboration, M. Derrick {\it et al.}, Phys. Lett. {\bf B316}
(1993) 412; Z. Phys. {\bf C65} (1995) 379.
\medskip

\item{[40]} M. Gluck, E. Reya and A. Vogt, Z. Phys. {\bf C53} (1992) 127;

\item{} J. G. Morfin and Wu-Ki Tung, Z. Phys. {\bf C52} (1991) 13.
\medskip

\item{[41]} V. Barone, M. Genovese, N. N. Nikolaev, E. Predazzi and
B. G. Zakharov. Z. Phys. {\bf C58} (1993) 541; Int. J. Mod. Phys. {\bf A8}
(1993) 2779.
\medskip

\item{[42]} A. Levy, {\it The energy behviour of real and virtual photon-proton
cross section}, DESY 85-003, talk at the fifth Gentner Symposium on Physics,
Dresden, 16-21, October, 1994.
\medskip

\item{[43]} R. D. Ball and S. Forte, {\it The rise de $F_{2}^{p}$ at HERA},
CERN-TH 7421/94, talk given at {\it The Heart of the Matter}, Blois, June 1995;
Phys. Lett. {\bf B336} (1994) 77;

\item{} F. Paccanoni, {\it Note on DGLAP evolution equation}, in Strong
Interaction at Large Distances, Hadronic Press, 1995, Ed. L. L. Jenkovszky.
\medskip

\item{[44]} a) M. Greco, Nucl. Phys. {\bf B63} (1973) 398;

\item{} b) A. Donnachie and P. V. Landshoff, Z. Phys. {\bf C61} (1994) 139;
\medskip

\item{} c) H. Abramovitz, E. M. Levin, A. Levy and U. Maor, Phys. Lett.
{\bf B269} (1991) 465.

\item{[45]} M. Bertini, P. Desgrolard, M. Giffon, L. L. Jenkovszky and
F. Paccanoni, {\it Phenomenological Analysis of the Small-$x$ Proton Structure
Function}, in Strong Interaction at Large Distances, Hadronic Press, 1995,
Ed. L.L. Jenkovszky.
\medskip

\item{[46]} A. Capella, A. Kaidalov, C. Merino and J. Tr\^an Thanh V\^an,
Phys. Lett. {\bf B337} (1994) 358.
\medskip

\item{[47]} M. Bertini, M. Giffon and  E. Predazzi, Phys. Lett. {\bf B349}
(1995) 561.
\medskip

\item{[48]} P. Desgrolard, M. Giffon and E. Predazzi, Z. Phys. {\bf C63}
(1994), 241.
\medskip

\item{[49]} M. Bertini, {\it Structure Functions from Photoproduction Process
to Deep-Inelastic Scattering}, in Strong Interaction at Large Distances,
Hadronic Press, 1995, Ed. L. L. Jenkovszky.
\medskip

\item{[50]} A. D. Martin, W. J. Stirling and R. G. Robert, Phys. Lett.
{\bf B354} (1995) 155.
\medskip

\item{[51]} G. Ingelman and P. Schlein, Phys. Lett. {\bf B152} (1985) 256.

\item{} G. Ingelman, Nucl. Phys. {\bf B} (Proc Suppl.) {\bf 18C} (1990) 172;
J. Phys.{\bf G 19}, {\it Workshop on HERA- the new Frontier for QCD}, (1994),
p.1631.
\medskip

\item{[52]} a) H1 Collaboration, T. Ahmed {\it et al}, {\it First Measurement
of
the Deep-Inelastic Structure of the Proton Diffraction}, preprint DESY 95-36,
(February 1995);

\item{} b) ZEUS Collaboration, M. Derrick {\it et al.}, Phys. Lett.
{\bf B 315} (1993) 481, Phys. Lett. {\bf B 332} (1994) 228.
\medskip

\item{[53]} a) CDF Collaboration, A. Bhatti, Proc. of the 2$^nd$ Workshop of
Small-$x$ and {\it Diffractive Physics at the Tevatron}, Fermilab, Sept.22-24,
 1994;

\item{} b) CDF Collaboration, F. Abe {\it et al}, Phys. Rev. Lett. {\bf 74}
(1995) 855;

\item{} c) DO Collaboration, G. Blazey {\it QCD at D0 and CDF}, Talk presented
at the {\it Les Rencontres de Physiques de la Vall\'ee D'Aoste, Results and
Perspectives in Particles Physics}, La Thuile Aosta, March 5-11, 1995.
\medskip

\item{[54]} J. D. Bjorken, Phys. Rev. {\bf D47} (1993) 101.
\medskip

\item{[55]} R. Fiore, L. L. Jenkovszky, and F. Paccanoni,
{\it Diffractive Deep-Inelastic Scattering}, DU-UNICAL/94-22 preprint,
Phys. Rev. {\bf D} in press.
\medskip

\item{[56]} a) A. Donnachie and P. V. Landshoff, Phys. Lett. {\bf B191} (1987)
309; Nucl. Phys. {\bf B303} (1988) 634;

\item{} b) A. Capella, A. Kaidalov, C. Merino and J. Tr\^an Thanh V\^an, Phys.
Lett.{\bf B343} (1995) 343;

\item{} c) J. C. Collins, J. Huston, J. Pumplin, H. Weert and J. J. Whitmore,
Phys. Rev. {\bf D51} (1995) 3182;

\item{} d) K. Goulianos, {\it The Structure of the Pomeron}, preprint
RU 95/E-26 (May 15), 1995.
\medskip
\vfill\eject

\centerline{\bf Figure captions}
\vskip 1 true cm

\noindent
{\bf Fig. 2.1}. Diagram for elastic hadron scattering with a single Regge pole
exchange, $s=(p_1+p_2)^2$ and $t=(p_1-p_2)^2$ are the Mandelstam variables,
$g_1(t), \ g_2(t)$ are the vertex (residue) functions and $s^{\alpha(t)}$ is
the Regge propagator.
\medskip

\noindent
{\bf Fig. 2.2.} Spin of the particles $J=\hbox{Re}\ \alpha(t)$ plotted versus
their squared masses $t=m^2$ (Chew-Frautschi plot). The function $\alpha(t)$ is
called a Regge trajectory; it realizes the analytic continuation from the
discrete and positive values of the spin of a particle to a continuum
$t\in ]-\infty , +\infty [.$
\medskip

\noindent
{\bf Fig. 2.3}. By unitarity, a simple Regge pole exchange may be related to
the
multiperipheral mechanism of particle production.
\medskip

\noindent
{\bf Fig. 2.4}. Data on the high energy behavior of the hadronic and
$\gamma p$ total cross sections [8 a)]. The curves in $\gamma p$ correspond
to model
fits  with a supercritical Pomeron (with $\alpha(0)-1=0.08$) [8 b)] (dotted
line), Froissart saturation (full line), and dipole Pomeron [8 c)]
(broken line).
\medskip

\noindent
{\bf Fig. 2.5}. Single (a) and double (b) diffraction dissociation are also
mediated by a Pomeron exchange.
\medskip

\noindent
{\bf Fig. 2.6}. Multi-Pomeron exchange diagram.
\medskip

\noindent
{\bf Fig. 2.7}. Pomeron trajectory fitted to the elastic scattering data with a
candidate glueball on it.
\medskip

\noindent
{\bf Fig. 3.1}. Two-gluon exchange is the "Born term" Pomeron [11, 12] in QCD.
\medskip

\noindent
{\bf Fig. 3.2}. Differential cross section for $pp$ scattering at
$\sqrt s=19.42$ GeV calculated in Ref. [24].
\medskip

\noindent
{\bf Fig. 3.3}. $\bar p p$ total cross section calculated in Ref. [24]
and fitted to the data.
\medskip

\noindent
{\bf Fig. 4.1}. Diagram for Deep Inelastic lepton-hadron Scattering (DIS),
mediated by one-boson exchange.
\medskip

\noindent
{\bf Fig. 4.2}. By unitarity, DIS is related to the imaginary part of the
forward virtual Compton elastic scattering amplitude.
\medskip

\noindent
{\bf Fig. 4.3}  Semi-exclusive DIS with a proton in the final state and a
vector hadronic system $X(J^{P,C}=1^{-,-})$ produced.
\medskip

\noindent
{\bf Fig. 4.4}. A pictorial presentation of the {\it "road map}" of various
regimes (and relevant approaches) to DIS.
\medskip

\noindent
{\bf Fig. 4.5}. Kinematical boundary of the Regge domain. Regge behavior is
expected to hold right and upwards from the curve indicated in the figure.
\medskip

\noindent
{\bf Fig. 4.6}. Calculated behavior of the SF in the Regge-type model of Ref.
[31].
\medskip

\noindent
{\bf Fig. 4.7}. Behavior of : (a) the input SF (with the assumed [34-36] Regge
behavior of the NMC data at $Q^2=5$ GeV$^2$), and  (b) its QCD evolution
according to the GLAP equation, calculated in Ref. [33] and with the HERA
data [38,39] superimposed.
\medskip

\noindent
{\bf Fig. 4.8}. $Q^2$-dependence of the "effective" Pomeron intercept from
Ref. [42].
\medskip

\noindent
{\bf Fig. 4.9 }. Small-$Q^2$ fit in a model [45] satisfying the $Q^2$ limit.
Note that models of that type fail to follow the rapid rise of the SF at
large $Q^2$, typical of the HERA data.
\medskip

\noindent
{\bf Fig. 4.10}. Calculated and predicted glue distribution in Ref. [41].
\medskip

\noindent
{\bf Fig. 4.11}. Behavior of the SF calculated in a model [46] combining small-
$Q^2$ Regge behavior (with a $Q^2$-dependent Pomeron intercept) and large-
$Q^2$ evolution.
\medskip

\noindent
{\bf Fig. 4.12}. Small-$x$ structure function $F_2^p$ from H1 data
(triangulated
dots) and Zeus data (closed points and stars) plotted as function of $x$ at
fixed $Q^2$ compared with the fit of Eq. (4.23). Only data with $x \le
5.10^{-3}$ have been used in the fit [47].
\medskip

\noindent
{\bf Fig. 4.13}. a) Structure function with the same data of {\bf Fig. 4.12}
plotted as a function of $x$ at $Q^2$ fixed and b) as a function of $Q^2$ at
$x$ fixed. The solid line is obtained with Eq. (4.23) where a factor
$(1-x)^{\beta}$ is included, $\beta$ being given by Eq. (4.26). Only data
with $x \le 5.10^{-3}$ have been used in the fit.
\medskip

\noindent
{\bf Fig. 4.14}. Fits to the data from a model [49] based on a two-component
model for the Pomeron.
\medskip

\noindent
{\bf Fig. 5.1}. Large rapidity gap events as first seen at HERA.
\medskip

\noindent
{\bf Fig. 5.2}. The Pomeron in DIS at large $Q^2$. The object (Pomeron)
remains the
same as in "soft" scattering, but its vertex becomes deeply inelastic. The
rapidity $ln W^2$ in this case is additive because  $W^2=Q^2({1\over x}-1)
+m^2\approx Q^2({1 \over x}-1)$ at large $W^2$.
\medskip

\noindent
{\bf Fig. 5.3}. The ratio of the diffractive to all deep inelastic events,
calculated in Ref. [55] and compared with the HERA data.
\end